\documentclass[prl,twocolumn,showpacs]{revtex4}
\usepackage{amsfonts,amsmath,mathrsfs,epsfig,amsbsy,bm,verbatim}

\begin{document}

\title{Majorana resonances and how to avoid them in periodic topological superconductor-nanowire structures}
\author{Jay D. Sau$^1$}
\thanks{Present address: Department of Physics, Harvard University, Cambridge, Massachusetts.}
\author{Chien Hung Lin$^{1}$}
\author{Hoi-Yin Hui$^{1}$}
\author{S. Das Sarma$^1$}
\affiliation{$^1$Condensed Matter Theory Center and Joint Quantum Institute, Department
of Physics, University of Maryland, College Park, Maryland 20742-4111, USA}

\begin{abstract}
Semiconducting nanowires in proximity to superconductors are  
promising experimental systems for Majorana
fermions which may ultimately be used as building blocks for topological quantum
computers.
A serious challenge in the experimental realization of the Majorana
 fermion in these semiconductor-superconductor nanowire
 structures is tuning the semiconductor chemical potential in close proximity to the
 metallic superconductor. 
We show that, presently realizable structures in experiments with tunable chemical potential 
 lead to \textit{Majorana resonances}, which are interesting in 
their own right, but do not manifest non-Abelian statistics. This poses a central challenge to the field.
We show how to overcome this challenge, thus resolving a crucial barrier
 to the solid state realization of a topological system containing the
 Majorana fermion. We propose  a new 
topological superconducting array structure where introducing 
the superconducting proximity effect from adjacent nanowires 
generates Majorana fermions with non-Abelian statistics. 
\end{abstract}

\maketitle

\paragraph{Introduction:}
Majorana fermions (MF) have been the subject of intense recent study,
both due to their fundamental interest as a new type of particle with non-Abelian statistics 
and their potential application in topological quantum computation
 (TQC)\cite{nayak_RevModPhys'08,Wilczek-3,levi,science}.
Topological superconductors are promising candidates for 
the practical solid state realization of  MFs
 \cite{sumanta,fu_prl'08,sau,long-PRB,alicea,roman,oreg}.
A simple topological superconducting (TS) 
systems supporting MFs, which has attracted serious 
experimental attention \cite{levi}, consists of a
 semiconductor nanowire in a  
magnetic field placed on an ordinary superconductor
 \cite{roman,oreg,long-PRB}.
The superconducting pair-potential
 is induced in the nanowire by proximity-effect from the $s$-wave
 superconductor (SC) in contact with the 
nanowire. It has been shown that such a nanowire can be driven into 
a TS phase with Majorana end modes 
  \cite{sau,roman,oreg}.
 The $s$-wave proximity effect on a
 InAs quantum wire, which also has a sizable SO coupling,
may have already been realized in experiments\cite{doh}.
 Therefore, it seems that a MF-carrying
 TS state in a semiconductor quantum wire may be within experimental
 reach. This has created a great deal of interest in the 
physics communtiy \cite{levi,science}.

\begin{figure}[tbp]
\includegraphics[width=1.0\columnwidth]{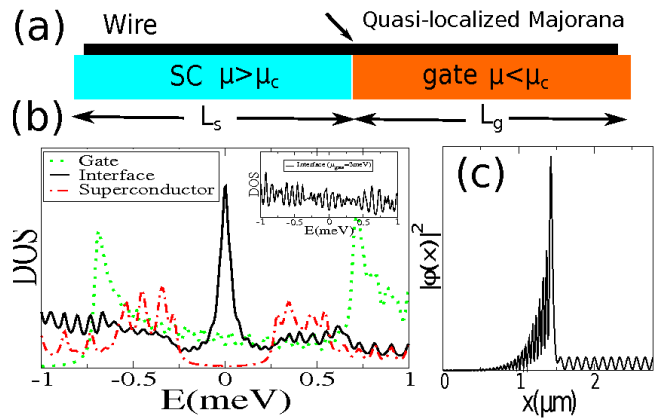}
\caption{(a) Geometry to observe zero-bias tunneling signature associated
with MFs in gated semiconductor nanowire (thick black line) 
 in the topological regime $(\protect\mu<\protect\mu_c)$.
Superconductivity is induced from the side of the gated nanowire region 
by the superconducting(blue box) layer on the left. (b) Tunneling 
density of states (DOS) of nanowire system shows broadened zero-bias tunneling 
peak at the interface for large Zeeman potential ($V_Z=0.75$ meV $> \Delta_0=0.5$ meV),
 gate chemical potential $\mu_g=0$ meV and spin-orbit energy $E_{SO}=\frac{m^*\alpha^2}{2}=100\,\mu$eV.
Inset shows that the near-zero-energy peak in the interface tunneling DOS goes away for 
a gate voltage in the non-topological regime $\mu_g=1.0$ meV.
 The chemical potential $\mu_s$ directly next to the
 SC is taken to be large ($\mu_s= 5$ meV $\gg \mu_c\approx 0.5$ meV).  (c) Zero-bias tunneling density 
of states shows quasilocalized states at interface that decays in
 SC and 
propagates in the gated region. }
\label{Fig1}
\end{figure}

In this Letter, we introduce the new theoretical concept of a 
Majorana resonance, in contrast to the well-studied Majorana bound 
state. Such Majorana resonances, which is an excitation that is closely 
related to MFs and is likely  occur 
in experimental set-ups designed to detect such MFs, do not manifest non-Abelian statistics. 
Here we consider  directly the SC-semiconductor nanowire structures being currently explored 
in experiments \cite{expts}, which have the schematic form shown in Fig.~\ref{Fig1}(a) and 
 are similar to the experiment that demonstrated the gate-tunable proximity-effect 
in InAs nanowires \cite{doh}. 
We show that such  structures \cite{doh, expts}, would produce Majorana resonances
 rather than the non-Abelian MFs.
 We then introduce a periodic structure which would
 lead to true MFs as required for TQC.

\paragraph{Nanowire Hamiltonian:}
The Majorana end modes together with other quasiparticle excitations of 
the superconducting semiconductor nanowire system (Fig. ~\ref{Fig1})\cite{long-PRB,roman,oreg} 
are described by a Bogoliubov-de Gennes(BdG) Hamiltonian which  can
be written as 
\begin{align}
& H_{BdG}=(-\frac{\hbar^2}{2 m^*} \partial _{y}^{2}-\mu (y))\tau _{z}+V_{z}\bm\sigma \cdot 
\hat{\bm B}+\imath \alpha \partial _{y}\hat{\bm\rho }\cdot \bm\sigma \tau
_{z}+\nonumber\\
&\Delta(y) \tau _{x}.\label{eq:1DBdG}
\end{align}
Here  the unit vector $\hat{\bm B}$ gives the direction of the effective
Zeeman field ($V_Z$), the unit vector $\hat{\bm\rho}$ characterizes the
spin-orbit coupling ($\alpha$) and $m^*$ and $\Delta(y)$ are, respectively, the effective mass and 
proximity-induced pairing potential. The $2\times 2$ Pauli matrices $\sigma_{x,y,z}$ and 
$\tau_{x,y,z}$ represent the spin and particle-hole degrees of freedom 
of the Bogoliubov quasiparticles. Such a semiconducting nanowire
can be tuned across a phase transition separating a TS and
non-topological superconducting (NTS) phase (i.e. an SC not containing any Majorana) simply by tuning
 either the chemical
potentual $\mu$ via a gate voltage or the Zeeman splitting $V_{Z}$ via an
in-plane magnetic field. In the TS phase, which is
reached by tuning $\mu$ and $V_{Z}$ to satisfy $|\mu|<\mu_c=\sqrt{V_{Z}^{2}-\Delta ^{2}}$
 \cite{sau, roman, oreg, long-PRB}, such a nanowire supports a pair of Majorana
zero energy modes (i.e. the non-Abelian MFs) at each end. The MFs, by
 virtue of their  non-Abelian statistics, can be used for 
 TQC \cite{nayak_RevModPhys'08,alicea1, david,flensberg,hassler, universal}. 
In fact,  MFs have been shown to exist in nanowires 
which are not strictly one-dimensional and
 several bands are occupied
 \cite{wimmer,roman_tudor}. 

While the simplicity of the recent proposals for realizing MFs
 has attracted a significant experimental and theoretical effort, several experimental hurdles 
towards realization of the MFs remain. One of the key challenges is the 
  requirement of 
 control of the chemical potential $\mu$ which
 has been realized in free-standing wires by the application of a gate voltage. 
Nanowires which form ohmic contacts with metals, instead of forming Schottky barriers,
 have a large density of electrons when in contact with the metal.  Gating of 
such  metallic nanowire segments that are directly in physical contact with an SC is ineffective  
because of strong electrostatic screening both from electrons in the underlying SC and the nanowire.
 Therefore, experimental attempts \cite{expts,doh} for the solid state realization
 of TS systems induce superconductivity in gated   segments of nanowire by the proximity effect
 from adjacent superconducting nanowire segments (see Fig.~\ref{Fig1}(a)) that are directly above 
the SC.
In this paper,  that proposed signatures of MFs are qualitatively weakened by 
such experimentally reasonable adaptations of the original 
proposals allowing suitable gating of the nanowire. 
In fact, the current experimental structures lead to Majorana resonance states as 
the end modes in the wires. These resonances are propagating states 
in the semiconductor sides of the nanowires, and as such, are not 
bound states with non-Abelian properties!
 We  propose a new periodic structure which should be ideal 
for the experimental MF realization.

 The recently proposed topological insulator (TI) nanowires,~\cite{franz} if realized in the absence of bulk dopants,
 can exist in a TS state over a large range of chemical potential $\mu$ and might avoid the generic complications 
discussed in this paper. However, it is likely that the discussion in this letter is relevant to the TI systems as well. 
The presence of bulk carries in the present experimental realization of TI nanowires, can be expected to lead to Majorana resonances as 
well. Moreover, the chemical potential $\mu$ needs to be reduced substantially for TI nanowires contacted to an SC  
so that the bulk states are not populated and more stringently, the mini-gap states are separated from the MFs by a 
substantial gap of $\Delta^2/\mu$~\cite{fu_prl'08,robustness}.

\paragraph{Zero-bias anomaly:}
The MFs that are predicted to occur in the TS phase of the 
semiconducting nanowire described by Eq.~\ref{eq:1DBdG} are associated with zero-energy
 end modes. These zero-energy modes can in principle be
 detected as a zero-bias conductance peak in the tunneling current 
at the end of the wire\cite{law,long-PRB}. This proposal constitutes one of
 the simplest signatures to test the existence of MFs in SCs.
 This tunneling experiment may be realized 
without gating the nanowire directly above the SC by using the set-up shown in
 Fig.~\ref{Fig1}(a) where  superconductivity is induced on a gated nanowire by the 1D 
proximity-effect from the SC on the left. A zero energy Majorana mode is expected to occur at the interface of
 the SC and gated region,  if the right half of the nanowire is gated to 
be effectively in the TS phase (i.e. $\Delta^2+\mu^2<V_{Z}^2$) while 
the left half of the nanowire above the SC is in the NTS phase. Since, strictly speaking, the pairing 
potential $\Delta$ vanishes in the gated segment of the wire, 
 we provide an alternative explanation for the Majorana resonances below. 

\paragraph{Majorana resonances:}
To  quantitatively verify whether such a zero-bias peak exists
 in the set-up in Fig.~\ref{Fig1}(a) we calculate the tunneling
 density of states for the BdG Hamiltonian
 of a nanowire (Eq. \ref{eq:1DBdG}) with periodic boundary conditions and
 length $(L=L_s+L_g)$, such that the interval $0< x< L_s$ is in contact
 with a SC (length $L_s$) and $L_s< x<L_s+L_g$ is gated
 (length $L_g$). 
The proximity-induced superconducting order parameter $\Delta(x)$ is
 taken to be $\Delta(x)=\Delta_0$  for $0< x< L_s$ which is in direct contact with
 the SC \cite{robustness,long-PRB} and taken to be $\Delta(x)=0$ otherwise. The chemical potential in the gated region can be
 controlled such that $\mu(x)=\mu_g<\mu_c=V_Z$.
 The chemical potential 
  elsewhere  is taken to be $\mu(x)=\mu_s\gg\mu_c$  
since the electron density in the immediate vicinity of  the
 SC could be large. The junction of the superconducting 
and gated regions at $x=L_s$ is expected to be analogous to the 
interface of nanowires in the TS phase (since $\mu_g <\mu_c$) and an
NTS phase (since $\mu_s >\mu_c$) and thus is expected to 
support an MF \cite{roman,oreg}.

The tunneling conductance at a given position on the nanowire, $x$, and at 
a given bias voltage $V$ relative to the SC,
 is proportional to the local density of states
  $(\rho(x;V)=\sum_{\sigma, E_n\approx V}|u_{n,\sigma}(x)|^2)$ 
 in the weak tunneling regime\cite{long-PRB}. Here $E_n$ and $(u_{n,\sigma}(x),v_{n,\sigma}(x))$
 are the BdG eigenvalues and eigenstates respectively of the continuum BdG Hamiltonian 
in Eq. $\left( \ref{eq:1DBdG}\right)$ which is calculated using a 
lattice approximation.  The results for the tunneling calculations in the gated
 region $(L_s<x<L_s+L_g)$, the interface $(x=L_s)$ and the superconducting region $(0<x<L_s)$ 
are shown in Fig.~\ref{Fig1}(b). The tunneling density of states  at the junction
is peaked near zero-bias as predicted~\cite{long-PRB}
 for MFs in topological wires while the density of states in the 
superconducting region shows a characteristic superconducting gap
 and  that in the gated region shows a  
uniform density of states characteristic of a metal. The inset in Fig.~\ref{Fig1}(b) shows that 
the Majorana resonance disappears for larger values of $\mu_g$.

The density of states in Fig.~\ref{Fig1}(b) can be understood from the multi-channel picture of MFs in quasi-one dimensional wires 
\cite{wimmer,roman_tudor,kitaev}. For large magnetic fields, $V_Z>\Delta$, the wire above the superconductor in Fig.~\ref{Fig1}(a) behaves like a multi-channel $p$-wave superconductor.
Some of these channels get terminated at the interface by the gate voltage. The termination of an
 odd number of channels leads to a Majorana bound state \cite{kitaev}.

  The peak associated with the MF is broadened in this set-up 
compared to previous predictions (where thermal broadening , neglected in our work,
 is the dominant contribution). This is a consequence of hybridization of the MFs
with the gapless excitations in the  gated nanowire which is not directly in contact with the SC.
This is clear from Fig.~\ref{Fig1}(c), where the wave-function of the state is found to be
 propagating (i.e. not decaying) in the gapless gated segment of the 
nanowire. Therefore the broadening of the MF is entirely analogous to the broadened Fano resonance that occurs
 when  a localized impurity is in contact with a bulk metal. 
Thus the MF here is a resonance, not a bound state - this is indeed a Majorana resonance!

\begin{figure}[tbp]
\includegraphics[width=.8\columnwidth]{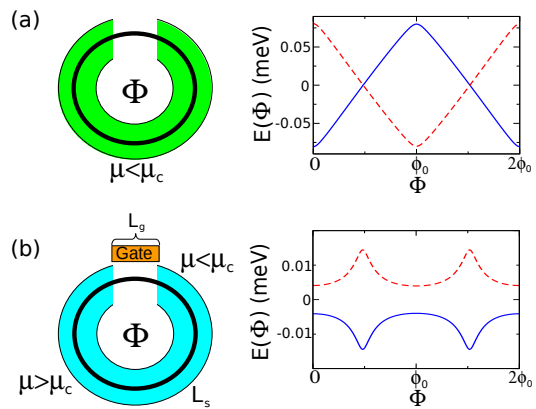}
\caption{(a) Junction in a ring topological superconductor structure with chemical
potential control over entire structure (such that $\mu\sim 0$ meV $< \mu_c$) shows a fractional 
Josephson effect. Flux dependence of  ABS energies and the corresponding 
Josephson current in junction show $2\Phi_0$ periodicity.
  (b) Experimental adaptations modify geometry so that $\mu_s>\mu_c$ in superconductor (length $L_s=1.5 \,\mu$m )
 with gate-induced chemical potential control only in junction 
$(\mu_g<\mu_c)$ (length $L_g= 600$ nm). ABS spectrum show a conventional Josephson effect in this  case despite tunneling signature of MFs in Fig.~\ref{Fig1}.}
\label{Fig2}
\end{figure}

\paragraph{Fractional Josephson effect:}
A definitive signature of MFs in TS nanowires is the fractional Josephson effect \cite{kitaev,yakovenko}.
The fractional Josephson effect not only probes the zero-energy character
of the MFs but is also a signature of its non-Abelian 
statistics \cite{kitaev}. It has been shown 
that a semiconducting nanowire in the TS phase $(\mu<\mu_c)$, 
placed on a superconducting ring in the geometry shown in
 Fig. \ref{Fig2}(a), would show a current versus phase relation which is 
$2\Phi_0=hc/e$ periodic in flux, $\Phi$, instead of the
 conventional $\Phi_0$ 
flux periodicity \cite{roman,oreg}.
 The Andreev bound state (ABS) spectrum ($E$ vs. $\Phi$) in the 
junction (gap in Fig.\ref{Fig2}(a)), shown in the left panel of
 Fig.\ref{Fig2}(a), determines the current-phase relation of the
 junction. The calculated function $E(\Phi)$ for the BdG Hamiltonian 
in Eq.~\ref{eq:1DBdG} for $\mu(x)<\mu_c$ and $\Delta(x)=\Delta_0$ shows 
the $2\Phi_0$ periodic $E(\Phi)$ in agreement with previous calculations 
\cite{roman,oreg}.

\paragraph{Absence of Josephson Fractionalization:}
Since it is difficult to apply a gate voltage to the nanowire directly above an SC,
 we consider a geometry (shown in Fig.~\ref{Fig2}(b)) so that $\mu(x)=\mu_g$ in the 
gated region inside the junction $(L_s<x<L_s+L_g)$, while $\mu(x)=\mu_s>\mu_c$
 in the nanowire segment in contact with the superconductor
 $(0<x<L_s)$. 
 The Josephson effect geometry involves a small junction in a long superconductor, so
 we  take the superconducting length $L_s\gg\xi$ where $\xi\sim 200$ nm
is the coherence length of the superconductor. 
Therefore the  ABS spectrum (shown in Fig.~\ref{Fig2}(b)) in the Josephson junction  
is periodic in the flux quantum $\Phi_0$. For the modified geometry in Fig.~\ref{Fig2}(b), 
the ABS spectrum does not cross $E=0$ at $\Phi=\Phi_0/2$.
 Therefore the corresponding current versus phase relation 
similarly only shows $\Phi_0$ periodicity.
 The Josephson effect fractionalization signature for MF  
is destroyed unless the chemical potential in contact with the
 superconductor $\mu_s$ can be tuned such that $\mu_s<\mu_c$. 
This seems to be a disaster since the Majorana resonance would not manifest any non-Abelian statistics!    

It is clear that simple structures where superconductivity is proximity-induced from 
adjacent nanowires cannot support true Majorana bound-state zero modes and 
 instead support only Majorana resonances which cannot be used for TQC.
These structures are thus at best 'half-topological-superconductors' (HTS) which can support Majorana resonance but not MF.
 In what follows, we propose a new TS structure that alternates between superconductor and gated 
regions (i.e. a periodic HTS system which is fundamentally inhomogeneous)
 that can realize true MFs with non-Abelian statistics. 

\begin{figure}[tbp]
\includegraphics[width=.8\columnwidth]{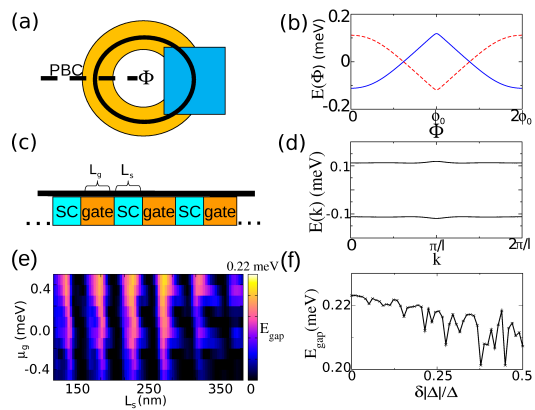}
\caption{(a) Short-superconductor geometry ($L_s \lesssim \xi\sim 200$ nm compared to 
Fig.~\ref{Fig2}(b)) to obtain fractional Josephson effect. (b) Corresponding ABS spectrum shows $2\Phi_0$ periodicity in current for ($L_s=100$ nm, $L_g=350$ nm and $\mu_g=0$ meV). (c) Periodic HTS structure corresponding to (a) is in TS phase because of fractional Josephson effect. (d) Band gap versus wave-vector 
in TS phase of periodic HTS structure. (e) Minimum band-gap in TS (with $Z=-1$ shown as bright) phase 
as a function of $L_s$ and $\mu_g$. The dark regions are either non-topological (i.e. $Z=1$) or have small band gaps.
(f) Minimum band-gap as a function 
of proximity-induced $\Delta$ magnitude fluctuations.  }
\label{Fig3}
\end{figure}

\paragraph{Periodic topological invariants:}
The identification of spatially inhomogeneous topological structures 
requires the use of topological invariants that are more complex
than the simple criterion $(\mu_g,\mu_s<\mu_c)$ \cite{akhmerov}.
For structures that are periodic in space, the topological character 
of the superconductor can be identified by calculating a Pfaffian
 topological invariant \cite{kitaev,parag,roman} 
\begin{equation}
Z=Sign(i^n Pf(H(k=0)\Lambda))Sign(i^n Pf(H(k=\pi)\Lambda))\label{eq:pfaffian}
\end{equation}
of the Bloch Hamiltonian $H(k)$ of the unit-cell of repetition.
Here $\Lambda=\sigma_y\tau_y$ and $k$ is the Bloch wave-vector 
along the periodic structure. Structures with $Z=-1$ 
are in a TS phase with MFs.
In fact, the fractional Josephson effect also occurs in a ring-geometry 
(shown in Fig.\ref{Fig2}) if the corresponding periodic system has 
 $Z=-1$. The periodic structure corresponding to the ring-geometry 
is obtained by cutting the ring structure at any point, 
straightening it out and then using this as the unit-cell of the 
periodic structure as shown in Fig.~\ref{Fig3}(a) and (c).
\paragraph{Fractional Josephson effect in rings with short superconductors:}
To realize a periodic structure which is topological it is necessary 
to find a system in a ring geometry that has a
 fractional Josephson effect.
The fractional Josephson effect is attributed to  
quasiparticle tunneling around the ring as opposed to Cooper pair
 tunneling responsible for the conventional Josephson 
effect\cite{yakovenko,kitaev}. Since Cooper pairs have electric charge 
$2e$, the energy levels associated with Cooper pairs in a ring are sensitive
 to flux in the ring with a period of $\frac{hc}{2 e}=\Phi_0$. In 
contrast, the flux periodicity of quasiparticle (with charge $e$) energy 
levels is twice as large ($\frac{hc}{e}=2\Phi_0$) . Therefore the 
 $2\Phi_0$ flux periodicity, which defines the fractional Josephson 
effect and is characteristic of TS nanowires, is 
natural for the persistent current in non-superconducting
 mesoscopic rings \cite{buttiker}.
 This suggests that the fractional Joesephson effect 
in Fig.~\ref{Fig2}(b) can be restored in a ring geometry with a 
 superconducting segment of length, $L_s$, that is smaller than the coherence
 length such that tunneling of quasiparticles around the ring is still allowed 
while maintaining a superconducting gap.
This can be seen from the plots of the ABS energies 
$E$ vs. $\Phi$ in Fig.~\ref{Fig3}(a). A small value of 
$L_s$ (lower plot) realizes a $2\Phi_0$ flux-periodicity
 characteristic of the 
fractional Josephson effect, while consistent with Fig.~\ref{Fig2} (b),
 large $L_s$ yields only a  conventional Josephson effect.
 As seen in Fig.~\ref{Fig3}(b) the presence of a superconducting segment,
 ensures that 
the structure is indeed gapped to quasiparticle excitations 
and is a \textit{bona fide} TS. The resulting one-dimensional TS can be tuned across a 
TS-NTS transition using a gate voltage and supports true zero-energy MFs at TS-NTS 
boundaries. Therefore all the schemes \cite{alicea1, david,flensberg,hassler, universal}
 proposed for TQC using TS nanowires \cite{roman,oreg} 
can be adapted to implement TQC in the proposed structure.

\paragraph{Phase-diagram and uneven structures:}
The dependence of the TS properties of the
 periodic HTS structure in Fig.~\ref{Fig3}(a) on the parameters $L_s,\mu_g$
is shown in Fig.~\ref{Fig3}(e). The color in the phase diagram
 in Fig.~\ref{Fig3}(e) only indicates the gap in the TS phase so that 
the dark regions are either nearly gapless or non-topological in 
nature. The phase diagram suggests that to obtain an optimal 
TS one must engineer structures with the 
appropriate values of $(L_s,\mu_g)$.
The oscillatory appearance of the topological phase in Fig.~\ref{Fig3}(e) 
is a result of the periodic $L_s$-dependence of the relative phases between the 
normal and Andreev transmission across the superconducting nanowire segments.  
Present experimental techniques allow the fabrication of structures 
with reasonably accurate feature length control. However the tunneling 
induced proximity effect, which depends on tunneling strength, can 
vary. Our calculation of the topological gap for an optimal 
parameter set(shown in Fig.~\ref{Fig3}(f)) indicates that the 
topological gap is robust to such disorder.  

\paragraph{Conclusion:} Experimental set-ups for realizing MFs and non-Abelian statistics in semiconductor nanowires 
are likely to be restricted to geometries where the chemical potential 
$\mu$ is only tunable in free-standing segments of the nanowire. 
We have found that this restriction qualitatively affects two of the 
simplest proposals for detecting MFs and forbids 
the realization of true non-Abelian MFs instead producing 
Majorana resonance modes. We resolve this 
problem by proposing a spatially inhomogenous periodic structure that 
can be engineered to support MFs at its ends. 
Our work thus resolves a key difficulty in the solid state realization of
 the MFs by providing an experimentally realizable architecture where the
 conflicting dichotomy of semiconductor gating and superconductor proximity effects
 could coexist harmoniously leading to non-Abelian particles potentially
 capable of carrying out TQC. We have introduced in this work the new 
concept of a Majorana resonance mode which is generically present in the 
currently studied superconducting-semiconductor structures \cite{sau,roman,oreg}
and topological insulator systems \cite{fu_prl'08,franz}.

We acknowledge useful discussions with C. M.  Marcus and L. P. Kouwenhoven.
This work was supported by DARPA-QuEST, JQI-NSF-PFC, and Microsoft-Q.

\end{document}